\documentclass[aps,showpacs,amsmath,amssymb,14pt]{revtex4}
\usepackage[latin9]{inputenc}
\usepackage{ifsym}
\usepackage{tipa}
\usepackage{amsmath}

\makeatletter

\newcommand{\lyxmathsym}[1]{\ifmmode\begingroup\def\b@ld{bold}
  \text{\ifx\math@version\b@ld\bfseries\fi#1}\endgroup\else#1\fi}

\@ifundefined{textcolor}{}
{%
 \definecolor{BLACK}{gray}{0}
 \definecolor{WHITE}{gray}{1}
 \definecolor{RED}{rgb}{1,0,0}
 \definecolor{GREEN}{rgb}{0,1,0}
 \definecolor{BLUE}{rgb}{0,0,1}
 \definecolor{CYAN}{cmyk}{1,0,0,0}
 \definecolor{MAGENTA}{cmyk}{0,1,0,0}
 \definecolor{YELLOW}{cmyk}{0,0,1,0}
}

\usepackage{latexsym}
\usepackage{bm}

\def\HollowBox #1#2{{\dimen0=#1 \advance\dimen0 by -#2
       \dimen1=#1 \advance\dimen1 by #2
        \vrule height #1 depth #2 width #2
        \vrule height 0pt depth #2 width #1
        \llap{\vrule height #1 depth -\dimen0 width \dimen1} 
       \hskip -#2
       \vrule height #1 depth #2 width #2}}

\makeatother

\begin{document}

\title{ On Exact Solutions and the Consistency of 3D Minimal Massive Gravity }

\author{Emel Altas, Bayram Tekin }

\email{btekin@metu.edu.tr}

\affiliation{Department of Physics,\\
 Middle East Technical University, 06800, Ankara, Turkey}

\date{\today}
\begin{abstract}

We show that all algebraic Type-$O$, Type-$N$ and Type-$D$ and some Kundt-Type  solutions of Topologically Massive Gravity are inherited by its holographically well-defined deformation, that is the recently found Minimal Massive Gravity.  This construction provides a large class of constant scalar curvature solutions to the theory. We also study the consistency of the field equations  both in the source-free and matter-coupled cases. Since the field equations of MMG do not come from a Lagrangian that depends on the metric and its derivatives only, it lacks the Bianchi identity valid for all non-singular metrics. But it turns out that for the solutions of the equations, Bianchi identity is satisfied. This  is a necessary condition  for the consistency of the classical field equations but not a sufficient one, since the the rank-two tensor equations are susceptible to double-divergence. We show that  for the source-free case the double-divergence of the field equations vanish for the solutions. In the matter-coupled case, we show that the double-divergence of the left-hand side and the right-hand side are equal to each other for the solutions of the theory. This construction completes the proof of the the consistency of the field equations.

\end{abstract}
\maketitle

\section{\label{intro} Introduction}

Recently \cite{Townsend1, Townsend2,Townsend3} a parity-non-invariant $2+1$ dimensional  massive gravity, dubbed as  Minimal Massive Gravity (MMG), with a single spin-2 massive degree
of freedom was found. MMG is a non-linear deformation of the Topologically Massive Gravity (TMG) \cite{Deser1} with the added property that the  bulk and the boundary unitarity
conflict of TMG in anti-de Sitter (AdS) space-time is resolved. Namely, positivity of the central charges of the two copies of the boundary Virasoro algebra and the unitarity of bulk gravitons
are simultaneously achieved in a certain parameter range in MMG, a desired property which TMG lacks. Hence MMG can have a viable, unitary boundary dual conformal field theory and so is amenable to the AdS/CFT duality. We shall briefly recapitulate the discussion below, but for a detailed account, the reader is referred to the recent activity in this topic \cite{Townsend1,Townsend2,Tekin,Alishahiha,Setare}. The matter-free field equation of the theory is {\it{defined}} as
\begin{equation} 
{\cal{E}}_{\mu\nu}\equiv{\bar{\sigma}}G_{\mu\nu}+\bar{\Lambda}_{0}\, g_{\mu\nu}+\frac{1}{\mu}C_{\mu\nu}+\frac{\gamma}{\mu^{2}}J_{\mu\nu}=0\,,\label{mmg_denk}
\end{equation}
with dimensionless parameters $\bar\sigma$, $\gamma$ and dimensionful ones $\mu$ and $\bar\Lambda_0$. We shall use the notation of the original work \cite{Townsend1} where the theory was introduced, except for slight notational differences such as in the definition of the anti-symmetric tensor below. The Cotton tensor is given in terms of the Schouten tensor as
\begin{equation}
C_{\mu\nu}=\eta_{\mu}\,^{\alpha\beta}\nabla_{\alpha}S_{\beta\nu},\hskip 1cm S_{\mu\nu}=R_{\mu\nu}-\frac{1}{4}g_{\mu\nu}R
\end{equation}
and the new ingredient is the $J$-tensor defined in terms of the products
of two Schouten tensors  as
\begin{equation}
J^{\mu\nu}\equiv \frac{1}{2}\eta^{\mu\rho\sigma}\eta^{\nu\alpha\beta}S_{\rho\alpha}S_{\sigma\beta}\, ,
\label{compact_J}
\end{equation}
where the totally anti-symmetric $\eta$-tensor reads in terms of the Levi-Civita
symbol $\eta^{\nu\rho\sigma}\equiv\epsilon^{\nu\rho\sigma}/\sqrt{-\det g}$
with $\epsilon_{012}=1$. In flat backgrounds, the $J$-tensor does
not contribute to the $O(h_{\mu\nu})$ fluctuations about the flat
metric $h_{\mu \nu} \equiv g_{\mu \nu} - \eta_{\mu \nu}$, hence the ''free'' particle properties of TMG (that is the
$\gamma=0$ case) is left intact. There is a single massive spin 2
particle (say with a +2 helicity) with a mass-square  $M_g^2=\mu^2 {\bar\sigma}$. On the other
hand, in AdS backgrounds which is the relevant background for holography, it is easy to show that the graviton's mass becomes
\begin{equation}
M_{g}^{2}=\mu^{2}\left(\bar\sigma+\frac{\gamma}{2\mu^{2}l^{2}}\right)^{2}+\frac{1}{l^{2}}
\label{massmmg}
\end{equation}
where $\text{\ensuremath{l}}>0$ is the AdS radius defined in terms of the effective cosmological constant as $\text{\ensuremath{l}}^{2}=-\frac{1}{\Lambda}$.
There are two possible values for $\Lambda$ that are determined by the quadratic equation coming from the field equation (\ref{mmg_denk})
\begin{equation} 
\bar\Lambda_{0}-\bar\sigma\Lambda+\frac{\gamma}{4\mu^{2}}\Lambda^{2}=0.
\end{equation}
Up to this point as far as the bulk excitations are concerned, the only apparent change in MMG from the TMG is the replacement 
$\bar\sigma\rightarrow\bar\sigma+\frac{\gamma}{2\mu^{2}\text{\ensuremath{l}}^{2}}$
in the  mass formula (\ref{massmmg}) plus the fact that there are two vacua in MMG  instead of the unique vacuum in TMG.  But,  once one computes the left and right central charges of the boundary Virasoro algebra as 
\begin{equation}
c_{L}=\frac{3\text{\ensuremath{l}}}{2G_{3}}\left(\bar\sigma+\frac{\gamma}{2\mu^{2}\text{\ensuremath{l}}^{2}}-\frac{1}{\mu\text{\ensuremath{l}}}\right), \hskip 1 cm  c_{R}=\frac{3\text{\ensuremath{l}}}{2G_{3}}\left(\bar\sigma+\frac{\gamma}{2\mu^{2}\text{\ensuremath{l}}^{2}}+\frac{1}{\mu\text{\ensuremath{l}}}\right)
\end{equation}
and the energies of bulk excitations
\begin{equation}
E_M = \frac{ M_{\mbox{g}}^2}{4 \pi G_3} \frac{1}{T} \int d^3 x \, \sqrt{-\bar{g}} \,\epsilon_\alpha\,^{0\mu} h^{\alpha \nu}\partial_t h^M_{\mu \nu}, \hskip 0.5 cm 
E_L= -\frac{ c_L} {6 \pi \ell } \frac{1}{T} \int d^3 x \, \sqrt{-\bar{g}} \,\bar{\nabla}^0  h^{\alpha \nu}_L \partial_t h^L_{\mu \nu},
\nonumber
\end{equation}
\begin{equation}
E_R= -\frac{ c_R} {6 \pi \ell } \frac{1}{T} \int d^3 x \, \sqrt{-\bar{g}} \,\bar{\nabla}^0  h^{\alpha \nu}_R \partial_t h^R_{\mu \nu},
\label{bulkexcitations}
\end{equation}
one realizes that the boundary unitarity constraints  $c_{L}>0$, $c_{R}>0$
and the bulk unitarity constraints $M_{g}^{2}\geq-\frac{1}{l^{2}}$, $ E \ge 0$ 
are compatible for a parameter region for non-zero $\gamma$. Remarkably, the $J$-tensor does its job: While keeping the bulk theory
intact, it makes the boundary conformal theory  unitarity opening up the possibility to find a dual CFT to the bulk gravity. That is certainly one way to define 
quantum gravity and it would be quite interesting to find the corresponding dual CFT of the MMG theory. All this is quite attractive but
the $J$-tensor has a slightly embarrassing defect : Its covariant divergence does not vanish for generic smooth metrics, instead one has 
\begin{equation}
\nabla_{\mu}J^{\mu\nu}=\eta^{\nu\rho\sigma}S_{\rho}{}^{\tau}C_{\sigma\tau},
\label{div_J_tensor}
\end{equation}
which means the MMG field equations do not obey the Bianchi identity and therefore cannot be obtained from an action with the metric being the only variable. (It can of course be derived from action(s) with auxiliary variables \cite{Townsend1,baykal} )
 But the state of affairs is not that bleak as the covariant divergence vanishes for metrics that are solutions to the full MMG equations. Therefore, one has an ''on-shell Bianchi identity".  This is good news but it actually makes it quite non-trivial to couple the theory to conserved energy-momentum tensors. This was achieved in \cite{Townsend2} in an intricate way. 

In the current work, we  shall consider two aspects of the MMG theory: The first being the consistency of  both the vacuum and matter-coupled  MMG equations and the second being the systematic  construction of new solutions to the vacuum field equations that are inherited from the TMG theory.   It will be clear below what we mean by the consistency of the field equations. As for the  solutions, there appeared several works on exact solutions of this theory in \cite{Townsend1,Townsend2,Alishahiha,Giribet,Arvanitakis}, but our work is different in the sense that we shall upgrade all the algebraic Types $O,N,D$ and some Kundt-Type solutions of TMG to be the solutions of MMG with simple modifications of the parameters. Our construction is inspired by several works \cite{aliev1,aliev2,aliev3} that used the solutions of TMG in finding the solutions of New Massive Gravity \cite{nmg}, a parity symmetric massive spin-2 theory,  and in finding  the solutions of all $f(Ricci)$ theories \cite{sisman}.

\section{Consistency of  Field Equations of MMG}

First, let us consider the source-free case, as mentioned above, consistency of the field equations requires that  the first divergence vanishes
\begin{equation}
\nabla_\mu {\cal{E}}^{\mu \nu} = 0, \hskip  0.5 cm \text{ on-shell}
\end{equation}
as  was already noted in \cite{Townsend1}, and worked out in that paper that it is on-shell valid: Namely, (\ref{div_J_tensor}) vanishes for the solutions.
Risking to be pedantic, let us note that, only {\it after} one takes the covariant divergence of the $J$-tensor,  one should insert the field equations, otherwise,  on-shell vanishing of the covariant divergence of the field equations would be a trivial statement for any theory since it would boil down to taking the derivative of zero. This statement also gives us the hint that one-shell vanishing of the first covariant derivative is necessary but not sufficient for the consistency of the field equations: Since one has a rank two tensor, one must show that the double divergence also vanishes on-shell. Namely one should show the following 
\begin{equation}
\nabla_\nu \nabla_\mu {\cal{E}}^{\mu \nu} =0,  \hskip  0.5 cm \text{ on-shell},
\end{equation}
which reduces to showing that the double divergence of the $J$-tensor vanishes on-shell. Let us show that this is indeed the case here. From (\ref{div_J_tensor}), one has 
\begin{equation}
\nabla_{\nu}\nabla_{\mu}J^{\mu\nu}=C_{\sigma\tau}C^{\sigma\tau}+\eta^{\nu\rho\sigma}S_{\rho}{}^{\tau}\nabla_{\nu}C_{\sigma\tau}.
\label{doub_div}
\end{equation}
We can compute the square of the Cotton tensor as
\begin{eqnarray}
C_{\mu\nu}C^{\mu\nu}&=&-\nabla^{\sigma}S^{\rho}\,_{\nu}\,\nabla_{\sigma}S_{\rho}\,^{\nu}+\nabla^{\rho}S^{\sigma}\,_{\nu}\,\nabla_{\sigma}S_{\rho}\,^{\nu} \\  \nonumber
&=&-\nabla^{\sigma}\Big(S^{\rho}\,_{\nu}\,\nabla_{\sigma}S_{\rho}\,^{\nu}\Big)+S^{\mu \nu}\Box S_{\mu \nu}+\nabla^{\rho}S^{\sigma}\,_{\nu}\,\nabla_{\sigma}S_{\rho}\,^{\nu}.
\end{eqnarray}
On the other hand, one can write the second term in (\ref{doub_div}) as
\begin{equation}
\eta^{\nu\rho\sigma}S_{\rho}{}^{\tau}\nabla_{\nu}C_{\sigma\tau}=-S^{\beta\tau}\Box S_{\beta\tau}+S^{\alpha\tau}\nabla_{\nu}\nabla_{\alpha}S^{\nu}\,_{\tau}.
\end{equation}
These expressions reduce the double-divergence of the $J$-tensor to an expression which is not zero for arbitrary metrics:
\begin{equation}
\nabla_{\nu}\nabla_{\mu}J^{\mu\nu}=-\frac{1}{2}\Box\Big(S^{\rho}\,_{\nu}\, S_{\rho}\,^{\nu}\Big)+\nabla^{\rho}\Big(S^{\sigma}\,_{\nu}\,\nabla_{\sigma}S_{\rho}\,^{\nu}\Big).
\end{equation}
Now, we need to use the field equations to show the vanishing of this expression. For this purpose, let us expand the compact form  (\ref{compact_J}) to get
\begin{equation}
J_{\mu\nu}  =-S_{\mu}^{\rho}S_{\rho\nu}+\frac{1}{2}g_{\mu\nu}S_{\rho\sigma}S^{\rho\sigma}+\frac{1}{4}S_{\mu\nu}R-\frac{1}{32}g_{\mu\nu}R^{2}.\label{J-ten}
\end{equation}
Using this form in (\ref {mmg_denk}), one arrives at 
\begin{equation}
\nabla^{\rho}\Big(S^{\sigma}\,_{\nu}\,\nabla_{\sigma}S_{\rho}\,^{\nu}\Big)=\frac{1}{2}\Box\Big(S^{\rho}\,_{\nu}\, S_{\rho}\,^{\nu}\Big), \hskip 1 cm \text{on-shell}.
\end{equation}
Hence the double divergence of the MMG field equations vanish for solutions of the theory as required for the consistency of these equations. 

 Let us now consider the consistency of  matter-coupled MMG equations \cite{Townsend2}  along the above discussion. (In this part, we use the definition of the $J$-tensor that  differs in a minus sign from the rest of our paper but it is consistent with the convention of \cite{Townsend2}). Because of the non-zero divergence of the $J$-tensor, as noted above,  coupling matter to MMG becomes quite a non-trivial issue. Given a covariantly conserved energy-momentum tensor  $\nabla_\mu T^{\mu \nu}=0 $, the field equations become 
\begin{equation}
\frac{1}{\mu}C_{\mu\nu}+\frac{\gamma}{\mu^{2}}J_{\mu\nu}+\eta G_{\mu\nu}=\Theta_{\mu\nu}(T),
\label{matter_coupled_mmg}
\end{equation}
where the source term reads
\begin{equation}
\Theta^{\mu\nu}(T)=\frac{b}{\gamma}T^{\mu\nu}+\frac{b^2}{\gamma \mu }\eta^{\mu\rho\sigma}\nabla_{\rho}\hat{T}{}_{\sigma}^{\nu}-\frac{b^2}{\mu^2} \eta^{\mu\rho\sigma}\eta^{\nu\lambda k}S_{\rho\lambda}\hat{T}_{\sigma k}
+\frac{b^{4}}{2\gamma \mu^{2}}\eta^{\mu\rho\sigma}\eta^{\nu\lambda k}\hat{T}_{\rho\lambda}\hat{T}_{\sigma k},
\label{source}
\end{equation}
where $b \equiv \gamma/(1+\gamma \eta)$ and $T_{\mu\nu}=-\bar{\rho}g_{\mu\nu}+\theta_{\mu\nu}$ and  with $\hat{T}_{\mu\nu}=T_{\mu\nu}-\frac{1}{2}g_{\mu\nu}T$ and $ \bar \rho $ is related to the bare cosmological constant of the theory.  For the consistency of the matter-coupled MMG, one should require the covariant divergence of the left-hand side and the right-hand side to be equal to each other when the field equations are used which was worked out to be the case in \cite{Townsend2}. Once again, this is necessary but not sufficient and one should also check the double divergence, which we calculate here.   The first divergence of (\ref{matter_coupled_mmg}) yields
\begin{equation}
\nabla_{\mu}\Theta^{\mu\nu}(T)=\frac{\gamma}{\mu^2} \nabla_{\mu} J^{\mu \nu} = \frac{\gamma}{\mu}\eta^{\nu\rho\sigma}S_{\rho}^{\lambda}\Theta_{\sigma\lambda}(T),
\label{div_theta}
\end{equation}
where the field equation was used in the second equality with the fact that products of  the form $(\eta S S S)^\nu $ vanish.  One can also explicitly compute this divergence from the definition of the source (\ref{source}) and get the same result. Next, we compute the double-divergence. From (\ref{div_theta}) one obtains 
\begin{equation}
\nabla_{\nu}\nabla_{\mu}\Theta^{\mu\nu}(T)=\frac{\gamma}{\mu}\eta^{\nu\rho\sigma}S_{\rho}^{\lambda}\nabla_{\nu}\Theta_{\sigma\lambda}(T)+\frac{\gamma}{\mu}\Theta_{\sigma\lambda}(T)C^{\sigma\lambda},
\end{equation}
in which we already made use of the field equations. Now, this result should match the direct computation of the double-divergence obtained from the definition of the source (\ref{source}):
 \begin{eqnarray}
\nabla_{\nu}\nabla_{\mu}\varTheta^{\mu\nu}&=& \frac{b^2}{\gamma \mu }\eta^{\mu\rho\sigma}\nabla_{\nu}\nabla_{\mu}\nabla_{\rho}\hat{T}_{\sigma}\thinspace^{\nu} 
-\frac{b^2}{ \mu^2 }\eta^{\mu\rho\sigma}\eta^{\nu\lambda k}\nabla_{\nu}\nabla_{\mu} \big (S_{\rho\lambda}\hat{T}_{\sigma k}\big )\nonumber \\
&&+\frac{b^4}{2 \gamma \mu^2 }\eta^{\mu\rho\sigma}\eta^{\nu\lambda k}\nabla_{\nu}\nabla_{\mu}\big ( \hat{T}_{\rho\lambda}\hat{T}_{\sigma k}\big ).
\end{eqnarray}
For the first term one can use the equality: $\eta^{\mu\rho\sigma}\nabla_{\mu}\nabla_{\rho}\hat{T}_{\sigma}\thinspace^{\nu}=\frac{1}{2}\eta^{\mu\rho\sigma}\left[\nabla_{\mu},\nabla_{\rho}\right]\hat{T}_{\sigma}\thinspace^{\nu}=\eta^{\nu\rho\sigma}S_{\rho}^{\lambda}\hat{T}_{\sigma\lambda}$ and for the last term one can use 
$\eta^{\mu\rho\sigma}\eta^{\nu\lambda k}\nabla_{\nu}\nabla_{\mu}\hat{T}_{\rho\lambda}\hat{T}_{\sigma k}=2\eta^{\mu\rho\sigma}\eta^{\nu\lambda k}\nabla_{\nu}(\hat{T}_{\rho\lambda}\nabla_{\mu}\hat{T}_{\sigma k})$.
Manipulations of this expression leads to the appearance of the Cotton tensor and one can use the field equations to get 
\begin{eqnarray}
\nabla_{\nu}\nabla_{\mu}\varTheta^{\mu\nu}&=& \frac{b^2}{\gamma \mu }\eta^{\nu\rho\sigma}S_{\rho}\thinspace^{\lambda}\nabla_{\nu}\hat{T}_{\sigma\lambda}+ \frac{b^2}{\gamma \mu }\eta^{\nu\rho\sigma}\hat{T}_{\sigma\lambda}\nabla_{\nu}S_{\rho}\thinspace^{\lambda} \nonumber \\
&&+\frac{\gamma}{\mu}C^{\sigma\lambda}\Theta_{\sigma\lambda}- \frac{b}{\mu }C^{\sigma\lambda}T_{\sigma\lambda}+\frac{\gamma b^2}{\mu^{3}}C^{\sigma\lambda}\eta_{\sigma}\thinspace^{\nu\rho}\eta_{\lambda}\thinspace^{\mu k}S_{\mu\nu}\hat{T}_{k\rho} \nonumber \\
&&-\frac{b^4}{2 \mu^3}C^{\sigma\lambda}\eta_{\sigma}\thinspace^{\nu\rho}\eta_{\lambda}\thinspace^{\mu k}\hat{T}_{\mu\nu}\hat{T}_{k\rho}-\frac{b^{2}}{\mu^{2}}\eta^{\mu\lambda k}\eta^{\nu\rho\sigma}S_{\rho\lambda}\nabla_{\nu}\nabla_{\mu}\hat{T}_{\sigma k} \nonumber \\
&&-\frac{b^{2}}{\mu^{2}}\eta^{\mu\rho\sigma}\eta^{\nu\lambda k}\nabla_{\nu}(\hat{T}_{\sigma k}\nabla_{\mu}S_{\rho\lambda})+\frac{b^{4}}{\gamma \mu^{2}}\eta^{\mu\rho\sigma}\eta^{\nu\lambda k}\nabla_{\nu}(\hat{T}_{\rho\lambda}\nabla_{\mu}\hat{T}_{\sigma k}).
\end{eqnarray}
This equation can be written in terms of the source-term by adding and subtracting terms to get
\begin{eqnarray}
\nabla_{\nu}\nabla_{\mu}\varTheta^{\mu\nu}&=&\frac{\gamma}{\mu}C^{\sigma\lambda}\Theta_{\sigma\lambda}+\frac{\gamma}{\mu}\eta^{\nu\rho\sigma}S_{\rho}\thinspace^{\lambda}\nabla_{\nu}\Theta_{\sigma\lambda} \nonumber \\
&&+\nabla_{\nu}\bigg(\frac{\gamma b^{2}}{\mu^{3}}\eta^{\lambda\mu k}\eta^{\nu\rho\sigma}\eta_{\sigma}\thinspace^{\alpha\beta}S_{\mu\alpha}\hat{T}_{k\beta}S_{\rho\lambda}-\frac{b^{4}}{2\mu^3}\eta^{\lambda\mu k}\eta^{\nu\rho\sigma}\eta_{\sigma}\thinspace^{\alpha\beta}\hat{T}_{\mu\alpha}\hat{T}_{k\beta}S_{\rho\lambda} \nonumber \\
&&+\frac{b^{4}}{\gamma \mu^{2}}\eta^{\lambda k\nu}\eta^{\mu\rho\sigma}\hat{T}_{\rho\lambda}\nabla_{\mu}\hat{T}_{\sigma k}-\frac{b}{\mu}\eta^{\nu\rho\sigma}\hat{T}_{\sigma\lambda}S_{\rho}\thinspace^{\lambda}+\frac{b^2}{\gamma \mu}\eta^{\nu\rho\sigma}S_{\rho}\thinspace^{\lambda}\hat{T}_{\sigma\lambda} \nonumber \\
&&-\frac{b^{2}}{\mu^{2}}\eta^{\mu\lambda k}\eta^{\nu\rho\sigma}\hat{T}_{\sigma k}\nabla_{\mu}S_{\rho\lambda}\bigg) .
\end{eqnarray}
The first line is what we want, therefore the other parts should vanish. To show that they add up to zero, we will have to use the field equation one more time. But first let us note that
\begin{equation}
\eta^{\lambda\mu k}\eta^{\nu\rho\sigma}\eta_{\sigma}\thinspace^{\alpha\beta}S_{\mu\alpha}\hat{T}_{k\beta}S_{\rho\lambda}=- \frac{1}{2}\eta^{\lambda\mu k}\eta^{\nu\rho\sigma}\eta_{\sigma}\thinspace^{\alpha\beta}S_{\mu\alpha}S_{k\beta}\hat{T}_{\rho\lambda}  =\eta^{\nu\rho\sigma}\hat{T}_{\rho\lambda}J_{\sigma}\thinspace^{\lambda}.
\end{equation}
Then one has
\begin{eqnarray}
\nabla_{\nu}\nabla_{\mu}\Theta^{\mu\nu}&=&\frac{\gamma}{\mu}C_{\sigma\lambda}\Theta^{\sigma\lambda}+\frac{\gamma}{\mu}\eta^{\nu\rho\sigma}S_{\rho}\thinspace^{\lambda}\Theta_{\sigma\lambda} \nonumber \\
&&+\nabla_{\nu}\bigg [\frac{b^{2}}{\mu}\eta^{\nu\rho\sigma}\hat{T}_{\rho}\thinspace^{\lambda}\left(\frac{\gamma}{\mu^{2}}J_{\sigma\lambda}+\frac{1}{\mu}C_{\sigma\lambda}\right)-\frac{b^{4}}{2\mu^{3}}\eta^{\nu\rho\sigma}\eta^{\lambda\mu k}\eta_{\sigma}\thinspace^{\alpha\beta}\hat{T}_{\beta k}\hat{T}_{\mu\alpha}S_{\rho\lambda}+\frac{b^{4}}{\gamma\mu^{2}}\eta^{\mu\rho\sigma}\eta^{\lambda k\nu}\hat{T}_{\rho\lambda}\nabla_{\mu}\hat{T}_{\sigma k} \nonumber \\
&&-\frac{b^{2}}{\gamma\mu}\eta^{\nu\rho\sigma}\hat{T}_{\rho\lambda}S_{\sigma}\thinspace^{\lambda}+\frac{b}{\mu}\eta^{\nu\rho\sigma}\hat{T}_{\rho\lambda}S_{\sigma}\thinspace^{\lambda}\bigg ].
\end{eqnarray}
Using the following relations one can show that the terms in the square bracket yield zero 
\begin{equation}
\eta^{\nu\rho\sigma}\eta^{\lambda\mu k}\eta_{\sigma}\thinspace^{\alpha\beta}\hat{T}_{\beta k}(\hat{T}_{\rho\lambda}S_{\mu\alpha}+\frac{1}{2}\hat{T}_{\mu\alpha}S_{\rho\lambda})=0, \hskip 1 cm \eta^{\rho\mu\sigma}\nabla_{\mu}\hat{T}_{\sigma k}=\eta_{k}\thinspace^{\mu\sigma}\nabla_{\mu}\hat{T}_{\sigma}\thinspace^{\rho}.
\end{equation}
This shows that the double divergence of the left hand-side and the right hand-side of the field equations are equal to each other on shell hence the equations are consistent.

\section{Exact Solutions of MMG}

In three dimensions, classification of space-times can be done either using the Cotton-tensor ( $C^\mu_\nu $)  ( analogous to the four dimensional Petrov classification )  or using the traceless Ricci tensor ($\widetilde{R}^\mu_\nu$ ) (analogous to the four dimensional Segre classification ).  For MMG, it is more convenient to use the Segre classification, but eventually for the solutions we shall consider these two classifications will be equivalent.  For Segre-Petrov classification, one needs the following  curvature invariants ( in addition to the scalar curvature $R$ )
\begin{equation}
I_1 \equiv \widetilde{R}^\mu_\nu \widetilde{R}^\nu_\mu, \hskip 0.5 cm I_2 \equiv \widetilde{R}^\mu_\nu \widetilde{R}^\nu_\rho  \widetilde{R}^\rho_\mu.
\end{equation}
For Types $O,N,III$ one has  $ I_1 = I_2 =0$ and for Types $D$ and $II$, one has $I_1^3 = 6 I_2^2$.   
To search for solutions of MMG, let us rewrite the source-free field equations as a trace part 
\begin{equation}
I_1-\frac{1}{24}R^{2}+\frac{\mu^{2}}{\gamma}\bar{\sigma}R-\frac{6\mu^{2}}{\gamma}\bar\Lambda_{0}=0,
\label{trace}
\end{equation}
and a traceless part
\begin{equation}
\frac{1}{\mu}C_{\mu\nu}+\bar{\sigma}\widetilde{R}_{\mu\nu}+\frac{\gamma}{\mu^{2}}\widetilde{J}_{\mu\nu}=0,
\label{traceless}
\end{equation}
where the traceless part of the $J$-tensor is given as
\begin{equation}
\widetilde{J}_{\mu\nu}=\widetilde{R}_{\mu \rho}\widetilde{R}^{\rho}_\nu-\frac{1}{3}g_{\mu\nu}I_1-\frac{1}{12}R \widetilde{R}_{\mu\nu}.
\label{tracless_J}
\end{equation}
We also need the TMG equations written in this form since the solutions of TMG will be upgraded to the solutions of MMG. The trace part of TMG equations simply says that $R = 6 \Lambda$, while the traceless part reads
\begin{equation}
\frac{1}{\mu}C_{\mu\nu}+\bar{\sigma}\widetilde{R}_{\mu\nu} = 0,
\label{tmg}
\end{equation}
to which we shall refer from now on.  Let us first consider all the solutions of MMG that satisfy 
\begin{equation}
\widetilde{J}_{\mu\nu} = 0,
\end{equation}
which boils down to all the solutions of TMG that has this property.   So, from (\ref{tracless_J}), one should set
\begin{equation}
\widetilde{R}_{\mu \rho}\widetilde{R}^{\rho}_\nu= \frac{1}{3}g_{\mu\nu}I_1+\frac{1}{12}R \widetilde{R}_{\mu\nu}.
\end{equation}
Contracting with one more traceless-Ricci tensor one arrives at
\begin{equation}
I_2 = \frac{1}{12} R I_1.
\end{equation}
Clearly  Type-$O$ solutions of TMG for which the canonical form of the traceless-Ricci tensor vanishes $\widetilde{R}_{\mu\nu}=0$, $\widetilde{J}_{\mu\nu} = 0$, hence all such solutions of TMG, which are locally Einstein spaces, also solve MMG. This statement might appear somewhat trivial but it actually has drastic implications. For example, the BTZ  black hole \cite{btz} which is a solution to cosmological Einstein's theory, survives to be a solution to TMG and also survives to be a solution to MMG. The solution is intact as a metric but its properties, such as its angular momentum and energy change for each theory.  See \cite{Deser_Tekin,Deser_Kanik}  for its mass and angular momentum properties in TMG and see \cite{Tekin} for those properties in MMG.

Next, we consider in a little more detail Type-$N$ and Type-$D$ solutions of MMG as inherited from TMG. As these solutions have been discussed in the literature before in the context of TMG \cite{nutku,gurses_94,pope1,pope2,aliev1,aliev2,aliev3,sisman}, we refer the reader to these works and especially to \cite{pope1} where these solutions are nicely compiled.

\subsection{Type-$N$ solutions of MMG}

The traceless Ricci tensor
\begin{equation}
\widetilde{R}_{\mu\nu}=R_{\mu\nu}-\frac{1}{3}g_{\mu\nu}R,
\end{equation}
for Type-$N$ space-times can be written as \cite{gurses}
\begin{equation}
\widetilde{R}_{\mu\nu}=\rho \xi_\mu \xi_\nu, 
\end{equation}
where $\rho$ is a scalar function which will not play a role and  $\xi_\mu$ is a null vector: $\text{\ensuremath{\xi}}_{\mu}\text{\ensuremath{\xi}}^{\mu}=0$.
For Type-$N$ space-times, since  $I_1=0$,  
from  (\ref{trace}) one concludes that the Ricci scalar is constant with two possible values
\begin{equation}
R_{\pm}=\frac{12\mu}{\gamma}  (\mu \bar\sigma \pm  m ), \hskip 1 cm m \equiv \sqrt{\mu^2\bar{\sigma}^2-\gamma \bar \Lambda_0}.
\end{equation}
Note that $m=0$ point is a special point ( "merger point"  \cite{Townsend2}) where two roots coalesce and needs seperate attention, which we note below.
The trace-free part of the $J$-tensor becomes
\begin{equation}
\widetilde{J}_{\mu\nu}=-\frac{1}{12}R \widetilde{R}_{\mu\nu},
\end{equation}
reducing the MMG field equations to 
\begin{equation}
\frac{1}{\mu}C_{\mu\nu}+\left(\bar{\sigma}-\frac{\gamma R }{12\mu^{2}}\right)\widetilde{R}_{\mu\nu}=0,
\end{equation}
which is nothing but the field equations of TMG (\ref{tmg})  with the simple  replacement of the parameters as 
\begin{equation}
\mu \bar\sigma \rightarrow  \mu\bar{\sigma}-\frac{\gamma R}{12\mu}.
\label{replacement0}
\end{equation}
Hence all Type-$N$ solutions of TMG solve MMG once this replacement is taken into account together with the values $ 6\Lambda = R_\pm$.  For $m=0$, observe that the traceless part of the MMG equation simply reduces to the vanishing of the Cotton tensor since $R = 12 \mu^2 \bar\sigma/\gamma$ at this point. So all such solutions are conformally flat spaces. 

Let us give an example of Type-$N$ solution \cite{olmez} which is locally equivalent to most Type-$N$ solutions of TMG, including the AdS-pp wave solutions \cite{pope1}
\begin{equation}
ds^2 =  d\rho^2 + e^{ 2 \rho/\ell } du dv + \big ( e^{( 1/\ell+ \mu\bar\sigma )\rho} f_1(u) + e^{ 2  \rho/\ell } f_2(u) + f_3(u) \big ) du^2
\label{ppp}
\end{equation}
with $R = -6 /\ell^2$, this is a solution to TMG for arbitrary functions $f_i(u)$. This solution also solves MMG after the replacement (\ref{replacement0}).   At the merger point, one can show that  (\ref{ppp}) becomes a conformally flat metric  but not an Einstein metric.

\subsection{Type-$D$ solutions}

Depending on the time-like or space-like nature of the eigenvectors of the traceless Ricci tensor, Type-$D$ solutions split into two as Type-$D_{t}$ and Type-$D_{s}$  \cite{gurses_94} and both types have the traceless Ricci tensor as
\begin{equation}
\widetilde{R}_{\mu\nu}=p\left(g_{\mu\nu}-\frac{3}{a}\text{\ensuremath{\xi}}_{\mu}\text{\ensuremath{\xi}}_{\nu}\right),
\end{equation}
where $\text{\ensuremath{\xi}}_{\mu}\text{\ensuremath{\xi}}^{\mu}\equiv a=\pm1$ and $p$ is a scalar function.  If the following equation is satisfied \cite{gurses_94}
\begin{equation}
\nabla_\mu \xi_\nu =\frac {\mu \bar\sigma}{3} \eta_{\mu \nu \rho} \xi^\rho,
\end{equation}
then TMG equation is solved with a constant $p$, as long as
\begin{equation}
 p = \frac{1}{9}\mu^2 \bar\sigma^2 + \frac{\bar\Lambda_0}{\bar\sigma}
\end{equation}
 For both Type-$D$ cases, one has 
\begin{equation}
\widetilde{R}_{\mu}\thinspace^{\rho}\widetilde{R}_{\nu\rho}=p^{2}\left(g_{\mu\nu}+\frac{3}{a }\xi_{\mu}\xi_{\nu}\right), \hskip 1 cm I_1=6p^{2}, \hskip 1 cm I_2 = -6 p^3.
\end{equation}
Then the  two roots of the trace-part of the field equation become
\begin{equation}
R_{\pm}=\frac{12\mu}{\gamma}  (\mu \bar\sigma \pm  M ), \hskip 1 cm M \equiv \sqrt{\mu^2\bar{\sigma}^2-\gamma \bar \Lambda_0+ \frac{\gamma^2 p^2 }{\mu^2}}.
\end{equation}
The new merger point is given by  $M=0$ which is generically satisfies by two possible $\Lambda_0$'s.  
On the other hand, the traceless part of the $J$-tensor becomes
\begin{equation}
\tilde{J}_{\mu\nu}=-\left(p+\frac{R}{12}\right)\widetilde{R}_{\mu\nu},
\end{equation}
reducing the MMG equation to the TMG equation as
\begin{equation}
\frac{1}{\mu}C_{\mu\nu}+\left(\bar{\sigma}-\frac{\gamma}{\mu^2}(p+\frac{R}{12})\right)\widetilde{R}_{\mu\nu}=0,
\end{equation}
which means all Type-$D$ solutions of TMG solve MMG once the following replacement is made
\begin{equation}
\mu \bar\sigma \rightarrow \mu \bar{\sigma}-\frac{\gamma}{\mu}(p+\frac{R}{12}) =   \mu \bar{\sigma}-\frac{\gamma}{\mu}\Big (\frac{1}{9}\mu^2 \bar\sigma^2 + \frac{\bar\Lambda_0}{\bar\sigma}+\frac{R}{12}\Big ).
\label{replacement}
\end{equation} 
Let us note that at the merger point and  for the  particular value of $\gamma  \bar \sigma = - 9$, the traceless part of the MMG equation becomes the sign-reversed version
\begin{equation}
\frac{1}{\mu}C_{\mu\nu}-\bar{\sigma}\widetilde{R}_{\mu\nu}=0,
\end{equation}
which in the TMG language refers to a change of helicity from +2 to -2 keeping the mass intact.

Let us now give two examples of such solutions. In \cite{pope1}, almost all Typ-$D$ solutions in the literature were shown to be locally equivalent to the time-like squashed $AdS_3$
\begin{equation}
ds^2 = \frac{ \lambda^2-4}{ 2 R}  \bigg ( - \lambda^2 (d\tau + \cosh \theta d \phi)^2 + d \theta^2 + \sinh^2 \theta d\phi^2 \bigg ),
\label{timelike}
\end{equation}
or the space-like squashed $AdS_3$
\begin{equation}
ds^2 =  \frac{ \lambda^2-4}{ 2 R}   \bigg ( - \cosh^2 \rho d \tau^2 + d \rho^2 + \lambda^2 ( dz +\sinh \rho d\tau)^2 \bigg ),
\end{equation}
with the squashing parameter $\lambda$, which for TMG reads
\begin{equation}
\lambda^2 = \frac{ 8 \bar\sigma^2 \mu^2}{  2 \bar\sigma^2 \mu^2 -9 R }.
\end{equation}
For these two solutions of TMG to also solve MMG, the squashing parameter changes according to  to the replacement recipe (\ref{replacement}) which we do not depict explicitly as it is clear.
For the sake of completeness, let us note that for  (\ref{timelike}), one finds the traceless part of the $J$-tensor as
\begin{equation}
\widetilde{J}_{\mu \nu} =- \frac{R}{4} \frac{(3 \lambda^2-4)}{(\lambda^2-4)} \widetilde{R}_{\mu \nu},
\end{equation}
while the square of the Cotton tensor reads
\begin{equation}
C_{\mu \nu} C^{\mu \nu} =  \frac{12 R^3 \lambda^2 ( \lambda^2-1)^2}{(\lambda^2-4)^3},
\end{equation}
for $\lambda= 1$ one has the round  $AdS_3$ metric and $ \lambda =  2$ corresponds to the flat space.

Finally, let us note that the following restricted version of the general Kundt solution of TMG reported in \cite{sezgin}  
\begin{equation}
ds^2 = 2 du dv+ \big(  \frac{1}{2}R-\frac{1}{9}\mu^2 \bar{\sigma}^2\big )  v^2 du^2 + \big (d\rho + \frac{2}{3} \mu \bar\sigma v du \big )^2 + du^2,
\end{equation}
also solves MMG since the traceless part of the $J$-tensor reads
\begin{equation}
\widetilde{J}_{\mu \nu} =- \frac{1}{4} \big ( R + \frac{4}{9}\mu^2 \bar\sigma^2 \big ) \widetilde{R}_{\mu \nu}.
\end{equation}

\section{Conclusions}

We have studied the consistency of the field equations of MMG by computing the on-shell vanishing of the double-divergence for both the matter-coupled and the source-free theories. 
We have also found a large class of solutions to MMG equations that are also solutions to the TMG equations. These solutions have constant scalar curvature and of Type-$O$ (here locally Einstein metrics such as the BTZ black hole), Type- $N$ solutions and Type-$D$ solutions in the Segre-Petrov classification.  We provided some explicit metrics that are called squashed $AdS_3$. Since MMG theory is free of bulk and boundary unitarity conflict, it is a good testing ground for ideas regarding holographic description of gravity. So we expect that the large classes of  solutions we have provided will be useful in this context. It would be interesting to see if there are some solutions with non-constant scalar curvature.

B.T. is partially supported by  T\"{U}B\.{I}TAK  grant 113F155. The Authors would like to thank M. Gurses and T.C. Sisman for extensive discussions about solutions of TMG and related theories.

\end{document}